\documentclass[prd,onecolumn,nofootinbib,showpacs,showkeys]{revtex4}
\usepackage{bm,amsmath,amssymb}
\newcommand\gothfamily{\usefont{U}{ygoth}{m}{n}}
\DeclareTextFontCommand{\textgoth}{\gothfamily}

\begin{document}

\title{A Maxwell field minimally coupled to torsion}
\author{Nikodem J. Pop{\l}awski}
\affiliation{Department of Physics, Indiana University, Swain Hall West, 727 East Third Street, Bloomington, Indiana 47405, USA}
\email{nipoplaw@indiana.edu}
\date{\today}

\begin{abstract}
We consider the Lagrangian density for a free Maxwell field, in which the electromagnetic field tensor minimally couples to the affine connection, in the Einstein-Cartan-Sciama-Kibble theory of gravity.
We derive the formulae for the torsion and electromagnetic field tensors in terms of the electromagnetic potential.
The divergence of the magnetic field does not vanish: the photon-torsion coupling acts like an effective magnetic monopole density.
Such a coupling, which breaks U(1) gauge invariance, is significant only at extremely high energies existing in the very early Universe or inside black holes.
It may, however, provide a mechanism for Dirac's quantization of electric charge.
\end{abstract}

\maketitle

The Einstein-Cartan-Sciama-Kibble (ECSK) theory of gravity \cite{KS} removes the constraint of general relativity that the torsion tensor (the antisymmetric part of the affine connection) be zero by promoting this tensor to a dynamical variable \cite{Hehl,rev,Niko}.
In this theory, the Lagrangian density for the gravitational field is proportional to the curvature scalar.
The field equations result from the stationarity of the total action under the variations of the metric, torsion and matter field tensors.
These equations naturally account for the intrinsic spin of Dirac fields.
Since such fields minimally couple to the affine connection, the torsion tensor in fermionic matter that composes all stars in the Universe is different from zero.
Macroscopic averaging of the conservation law for this spin tensor leads to the description of fermionic matter as a spin fluid \cite{NSH}.

The field equations of the ECSK gravity can be written as the general-relativistic Einstein equations in which the modified energy-momentum tensor has terms that are quadratic in the spin tensor \cite{KS,Hehl,rev,Niko}.
For the spin fluid, these terms generate gravitational repulsion, which is significant only at ultranuclear densities and prevents the formation of unphysical singularities \cite{avert}.
The singular big bang is thus replaced by a nonsingular bounce, before which the Universe was contracting \cite{Kuch,infl}.
Such a bounce naturally explains, without cosmic inflation, why the present Universe appears spatially flat, homogeneous, and isotropic \cite{infl}.
Torsion also modifies the classical Dirac equation by generating the cubic Hehl-Datta term \cite{HD}, which may be the source of the observed matter-antimatter imbalance and dark matter in the Universe \cite{bar}.
Furthermore, the gravitational interaction of condensing fermions arising from the Hehl-Datta term may be the source of dark energy \cite{dark}.

Since the covariant derivative $\nabla_i$ (with respect to the affine connection) of a spinor in the ECSK gravity contains the torsion tensor $S^i_{\phantom{i}jk}$, we would expect the covariant derivative of the electromagnetic potential $A_i$ to define the electromagnetic field tensor:
\begin{equation}
F_{jk}=\nabla_j A_k-\nabla_k A_j=\partial_j A_k-\partial_k A_j+2S^i_{\phantom{i}jk}A_i.
\label{Faraday}
\end{equation}
We use the notations of \cite{Niko}.
Such a tensor is minimally coupled to gravity, but it is not invariant under a local U(1) gauge transformation $A_i\rightarrow A_i+\partial_i f(x^j)$.
Basing electrodynamics on the conservation laws of electric charge and magnetic flux leads to the definition of the electromagnetic field tensor without any dependence on the metric or connection: $F_{jk}=\partial_j A_k-\partial_k A_j$ \cite{mini}.
Such a tensor is invariant under a U(1) gauge transformation.
The minimal coupling of electromagnetism and gravity is compatible with electromagnetic gauge invariance even if torsion is present, provided that a local gauge transformation is modified and the torsion tensor is a function of the gradient of a scalar \cite{HRRS}.
It is also possible to introduce nonminimal interaction of the electromagnetic potential and torsion that preserves local gauge invariance \cite{SG}.

U(1) gauge invariance and no photon-torsion coupling in Maxwell electrodynamics are related to the fact that a photon is massless.
Massive gauge bosons of the weak interaction can, however, be minimally coupled to the affine connection that contains torsion without restricting the structure of the torsion tensor and only at the cost of gauge symmetry breaking \cite{rev}.
According to the standard model of elementary particles, these vectors acquire mass through spontaneous symmetry breaking \cite{MS}.
Vector fields can also acquire mass in the ECSK gravity with additional terms containing torsion \cite{mass}.
Torsion may thus be involved in generating masses of elementary particles  and, in return, be coupled only to particles that are massive.

We consider a free Maxwell field in the ECSK gravity, assuming that it minimally couples to the affine connection.
The Lagrangian density for such a field is
\begin{equation}
\mathfrak{L}=-\frac{1}{16\pi}F_{jk}F^{jk}\sqrt{-g},
\label{Lagrange}
\end{equation}
where the electromagnetic field tensor $F_{jk}$ is given by (\ref{Faraday}).
The spin density $\mathfrak{S}_{ijk}$ satisfies $\mathfrak{S}_{ijk}=\tau_{[ij]k}$, where $\tau_{i}^{\phantom{i}jk}=2\frac{\delta\mathfrak{L}}{\delta S^{i}_{\phantom{i}jk}}$ \cite{rev,Niko}.
The Lagrangian density (\ref{Lagrange}) gives $\tau_{ijk}=-\frac{1}{4\pi}A_i F_{jk}\sqrt{-g}$ and
\begin{equation}
\mathfrak{S}_{ijk}=-\frac{1}{8\pi}(A_i F_{jk}-A_j F_{ik})\sqrt{-g}.
\label{spindensity}
\end{equation}
The Cartan equation relates the torsion tensor to the spin density according to $S_{ijk}-S^l_{\phantom{l}jl}g_{ik}+S^l_{\phantom{l}kl}g_{ij}=-\frac{\kappa}{2\sqrt{-g}}\mathfrak{S}_{jki}$ or $S^k_{\phantom{k}ij}=-\frac{\kappa}{2\sqrt{-g}}\bigl(\mathfrak{S}_{ij}^{\phantom{ij}k}+\delta^k_{[i}\mathfrak{S}_{j]l}^{\phantom{j]l}l}\bigr)$ \cite{KS,Hehl,rev,Niko}.
Substituting (\ref{spindensity}) into this equation gives the torsion tensor:
\begin{equation}
S^k_{\phantom{k}ij}=\frac{\kappa}{16\pi}A_{[i}F_{j]}^{\phantom{j]}k}.
\label{torsiontensor}
\end{equation}
The electromagnetic field tensor (\ref{Faraday}) therefore satisfies $F_{ij}=\partial_i A_j-\partial_j A_i+\frac{\kappa}{16\pi}A^k(A_i F_{jk}-A_j F_{ik})$, which can be written as
\begin{equation}
F_{kl}\biggl(\frac{1}{2}\delta^k_i \delta^l_j-\frac{1}{2}\delta^k_j \delta^l_i+\frac{\kappa}{16\pi}A_i A^k \delta^l_j-\frac{\kappa}{16\pi}A_j A^k \delta^l_i\biggr)=\partial_i A_j-\partial_j A_i=\nabla^{\{\}}_i A_j-\nabla^{\{\}}_j A_i,
\label{EMfield}
\end{equation}
where $\nabla^{\{\}}_i$ is the covariant derivative with respect to the Christoffel symbols.

Introducing an auxiliary tensor $H_{kl}$ satisfying $H_{[kl]}=F_{kl}$ turns (\ref{EMfield}) into $H_{kj}M^k_i=\nabla^{\{\}}_i A_j$, where $M^k_i=\delta^k_i+\frac{\kappa}{8\pi}A_i A^k$.
Accordingly, $H_{kj}=(M^{-1})^i_k\nabla^{\{\}}_i A_j$, where $(M^{-1})^i_j M^j_k=\delta^i_k$.
By checking the last relation one can show that $(M^{-1})^i_k=\delta^i_k-(\frac{\kappa}{8\pi}A^i A_k)/(1+\frac{\kappa}{8\pi}A_l A^l)$.
The electromagnetic field tensor (\ref{Faraday}) is therefore expressed in terms of the electromagnetic potential as
\begin{equation}
F_{kj}=\biggl(\delta^i_k-\frac{\frac{\kappa}{8\pi}A^i A_k}{1+\frac{\kappa}{8\pi}A_l A^l}\biggr)\nabla^{\{\}}_i A_j-\biggl(\delta^i_j-\frac{\frac{\kappa}{8\pi}A^i A_j}{1+\frac{\kappa}{8\pi}A_l A^l}\biggr)\nabla^{\{\}}_i A_k.
\label{EMtensor}
\end{equation}
Substituting this relation into (\ref{torsiontensor}) gives the torsion tensor in terms of the electromagnetic potential:
\begin{equation}
S_{kij}=\frac{\kappa}{32\pi}\biggl(A_i(\partial_j A_k-\partial_k A_j)-A_j(\partial_i A_k-\partial_k A_i)+\frac{\frac{\kappa}{8\pi}A_k A^m}{1+\frac{\kappa}{8\pi}A_l A^l}\bigl(A_i\nabla^{\{\}}_m A_j-A_j\nabla^{\{\}}_m A_i\bigr)\biggr).
\label{tensor}
\end{equation}
The corresponding modified Maxwell equations follow from substituting (\ref{EMtensor}) into the Lagrangian density (\ref{Lagrange}) and varying the corresponding total action with respect to $A_i$.
These equations, in the presence of Dirac sources that also couple to torsion and thus further modify $F_{ik}$, will be studied elsewhere.

If $\sqrt{\kappa}|A^i|\ll 1$ then (\ref{EMtensor}) reduces to the usual, gauge-invariant definition $F_{kj}=\partial_k A_j-\partial_j A_k$.
For the electrostatic field of a charge $e$ at a distance $r$ from it, in a gauge in which the electric potential vanishes at infinity, this condition  reads
\begin{equation}
r\gg r_e=\frac{\sqrt{G}e}{c^2},
\end{equation}
where $r_e$ is the Reissner-Nordstr\"{o}m length scale corresponding to this charge.
For an electron, $r_e$ is one order of magnitude smaller than the Planck length.
The deviation of $F_{ik}$ from $\partial_i A_k-\partial_k A_i$, which breaks U(1) gauge invariance, is therefore significant only at extremely high energies that can probe such lengths, existing in the very early Universe and inside black holes.
Even at the highest energies that are available at the LHC, such a deviation is negligible.
The electromagnetic field tensor (\ref{Faraday}) and the Lagrangian density (\ref{Lagrange}) are therefore effectively gauge invariant, provided that one begins with a natural (the Lorentz or Coulomb) gauge and that the four-gradient of a function $f(x^j)$ in a local U(1) gauge transformation $A_i\rightarrow A_i+\partial_i f(x^j)$ does not significantly exceed in magnitude the original electromagnetic potential.
An example of such a ``good'' transformation is the one used to derive the Darwin Lagrangian of a system of charges to terms of order $\frac{v^2}{c^2}$ \cite{LL}.
Implications regarding the mass of a photon associated with this restricted gauge invariance will be studied elsewhere.

In Maxwell electrodynamics, the relation $\nabla^{\{\}}_{[i}F_{jk]}=0$ contains the fact that the divergence of the magnetic field vanishes: magnetic charges (monopoles) do not exist.
If the electromagnetic potential minimally couples to the affine connection in the ECSK gravity, however, then such a relation does not hold.
Instead, the electromagnetic field tensor (\ref{Faraday}) satisfies
\begin{equation}
\nabla^{\{\}}_{[m}F_{kj]}=-\frac{\kappa}{4\pi}\nabla^{\{\}}_{[m}\biggl(\frac{A^i A_k\nabla^{\{\}}_{|i|}A_{j]}}{1+\frac{\kappa}{8\pi}A_l A^l}\biggr).
\label{anti}
\end{equation}
A nonvanishing value of $\nabla^{\{\}}_{[i}F_{jk]}$ can be associated with the magnetic four-current density $\mathfrak{j}^i_\textrm{m}$ (which is a spacetime pseudovector density) through $3\nabla^{\{\}}_{[m}F_{kj]}=\frac{4\pi}{c}\epsilon_{mkjn}\mathfrak{j}^n_\textrm{m}$.
The relation (\ref{anti}) leads to
\begin{equation}
\mathfrak{j}^n_\textrm{m}=\frac{\kappa c}{32\pi^2}\epsilon^{nmkj}\nabla^{\{\}}_m\biggl(\frac{A^i A_j\nabla^{\{\}}_i A_k}{1+\frac{\kappa}{8\pi}A_l A^l}\biggr).
\end{equation}
The corresponding magnetic charge density in the locally geodesic frame of reference is
\begin{equation}
\rho_\textrm{m}=\frac{\mathfrak{j}^0_\textrm{m}}{c}=-\frac{\kappa}{32\pi^2}\mbox{div}\biggl(\frac{{\bf A}\times(A^i\partial_i{\bf A})}{1+\frac{\kappa}{8\pi}A_k A^k}\biggr).
\end{equation}
The divergence of the magnetic field does not vanish: the photon-torsion coupling generates an effective magnetic charge density.
Such an effective magnetic monopole density may provide a mechanism for Dirac's quantization of electric charge \cite{Dir}, which would be another advantage of torsion and the ECSK theory of gravity.

\end{document}